\documentstyle[11pt,epsfig]{article}
\author{{Malihe Heydari-Fard}\thanks{E-mail:
heydarifard@qom.ac.ir} \thanks{E-mail:
m.heydarifard@mail.sbu.ac.ir}
\\{\small {{Department of Physics, The University of Qom, P.
O. Box 37155-1814, Qom, Iran}}}}
\begin{document}
\title{\bf Effect of bulk Lorentz violation on anisotropic brane cosmologies}
\maketitle 
\begin{abstract}
The effect of Lorentz invariance violation in cosmology has
attracted a considerable amount of attention. By using a dynamical
vector field assumed to point in the bulk direction, with Lorentz
invariance holding on the brane, we extend the notation of Lorentz
violation in four dimensions \cite{Jacobson} to a five-dimensional
brane-world. We obtain the general solution of the field equations
in an exact parametric form for Bianchi type I space-time, with
perfect fluid as a matter source. We show that the brane universe
evolves from an isotropic/anisotropic state to an isotropic de
Sitter inflationary phase at late time. The early time behavior of
anisotropic brane universe is largely dependent on the Lorentz
violating parameters $\beta_i, i = 1, 2, 3$ and the equation of
state of the matter, while its late time behavior is independent
of these parameters.
\vspace{5mm}
\vspace{0.5mm}
\end{abstract}

\section{Introduction}

Brane-world scenarios with a 3-brane identified with the
observable universe which is embedded in a higher-dimensional
space-time provide an alternative to the standard cosmology in
four-dimension \cite{RS, DGP1, DGP2}. A essential requirement on
these models is that they should reproduce the important
observational cosmological data, the age of the universe,
abundances of elements produced in primordial nucleosynthesis,
etc. One of the most successful of such higher-dimensional models
is that proposed by Randall and Sundrum whose the bulk has the
geometry of an AdS space admitting $Z_2$ symmetry \cite{RS}. They
were successful in explaining the hierarchy problem: the enormous
disparity between the strength of the fundamental forces. The
Randall Sundrum (RS) model has had a large impact on our
undersetting of the universe and has brought higher-dimensional
gravitational theories to the fore. In RS type models, all matter
and gauge fields live on the brane, while gravity can propagate
into the bulk. Using the Israel junction conditions \cite{Israel}
and the Gauss-Codazzi equations, one can obtain the gravitational
field equations on the brane, as employed by Shiromizu, Maeda and
Sasaki (SMS) \cite{SMS}. There are two important difference that
result from the effective four-dimensional gravitational equations
on the brane. The first one is quadratic energy-momentum tensor,
$\pi_{\mu\nu}$, which is important in a high-energy universe and
the second one is the projected Weyl tensor, ${\cal E}_{\mu\nu}$,
on the brane which is responsible for carrying on the brane the
contribution of the bulk gravitational field. The cosmological
evolution of such a brane universe has been
investigated and effects such as a quadratic density term in the
Friedmann equations have been found \cite{review1, review2,
review3}.

Lorentz symmetry is assumed to be the exact symmetry of nature
\cite{1}. However, many exotic theories such as canonical quantum
gravity and string theory suggest that Lorentz invariance may be
broken at high energies \cite{2,3}. String theory predicts that we
may live in a universe with non-commutative coordinate \cite{4}
leading to violation of Lorentz invariance as an explanation for
the astrophysical anomalies such as the missing
Greisen-Zatsepin-Kuzmin cutoff \cite{05, 5, 010, 10}. A method of
implementing the local Lorentz violation in a gravitational
setting is to consider the existence of a tensor field with a
non-vanishing expectation value and then couple it to gravity or
matter fields. The simplest example of this method is to consider
a single time-like vector field with fixed norm. A special case of
this theory as a mechanism for Lorentz invariance violation has
been introduced by Kostelecky and Samuel \cite{VA}. In a different
context, a theory of gravity with a fixed norm vector in order to
mimic the effects of dark matter has been proposed by Bekenstein
\cite{21}. Also, for the studies of vector fields in a
cosmological setting without the fixed norm see \cite{022, 22}.
This vector field picks out a preferred frame at each point in
space-time and any matter field coupled to this vector field will
experience a violation of local Lorentz invariance \cite{new}.
However, in curved space-time there is no natural generalization
of the notion of a constant vector field ($\nabla_{\mu} N^{\nu} =
0$ generically has no solutions); therefore we must allow that the
vector field to have dynamic and fix the norm of it by choosing an appropriate action
for the field.

The effects of Lorentz violation have been also studied within the
context of the brane-world scenarios \cite{b}. In these models,
the space-time violates four-dimensional Lorentz invariance
globally and leads to apparent violations of Lorentz invariance
from the brane observer's point of view due to bulk gravity
effects. These effects are restricted to the gravity sector of the
effective theory while the measured Lorentz invariance of particle
physics remains unaffected in these models \cite{29}. In a similar
way, Lorentz invariance violation has had to shed some light on
the possibility of signals travelling along the extra dimension
outside our universe \cite{31}. In a different method a brane-world toy model in
an inflating five-dimensional brane-world scenario has been introduced with violation
of four-dimensional Lorentz invariance at an energy scale $k$
\cite{Rubakov}. In \cite{Ahmadi}, the authors have studied Lorentz violation in a
brane-world model by considering a vector field normal to the
brane along the extra dimension. They showed that local Lorentz
violation in the bulk allows for the construction of models in
which the vacuum energy, the gravitational coupling and the
cosmological term on the brane are variable. There also exist a
relation between the maximal velocity in the bulk and the speed of
light on the brane \cite{Jalalzadeh}. The main aim of this paper
is to see how the shear parameter behaves in such brane-world
scenario and whether this model isotropises at late time. To
achieve this goal we study the local Lorentz invariance violation
in the same setting and investigate the behavior of the
observationally important parameters like shear, anisotropy, and
the deceleration parameter in a Bianchi type I geometry.

\section{The setup}

Let us start by presenting the model used in our calculations. We
only state the results and refer the reader to \cite{Ahmadi} for a
detailed derivation of these results.

As mentioned above, the brane-world model we consider here
includes Lorentz violation in the bulk space along the extra
dimension by generalizing the theory suggested by Jacobson and
Mattingly \cite{Jacobson, Kostelecky}, where the quartic
self-interaction term $(N^A\nabla_A N^B)(N^C\nabla_CN_B)$ has been
ignored \cite{32}. We assume that $N^A$ is a vector field along
the extra dimension which is making the associated frame a
preferred one. We also consider that the theory which is consists
of the vector field $N^A$ minimally coupled to gravity with the
following action
\begin{equation}
{\cal S} = \int d^5x
\sqrt{-g^{(5)}}\left[\frac{1}{2\kappa_5^2}\left(R^{(5)} + {\cal
L}_N \right)+ {\cal L}_m\right],
\end{equation}
where $\kappa_5^2 = 8\pi G_5$, $R^{(5)}$ is the five-dimensional
Ricci scalar, ${\cal L}_N$ is the vector field Lagrangian density
and ${\cal L}_m$ is the Lagrangian density for the matter fields.
We take ${\cal N}^A$ as a dynamical field for preserving the
general covariance. The Lagrangian density for the vector field is
given by
\begin{equation}\label{3a}
{\cal L}_{N} = K^{AB}_{\,\,\,\,\,\, CD} \nabla_A N^C\nabla_B N^D
+\lambda(N_AN^A-\epsilon),
\end{equation}
where $\lambda$ is a Lagrange multiplier and
\begin{equation}\label{}
K^{AB}_{\,\,\,\,\,\, CD} = -\beta_1
g^{AB}g_{CD}-\beta_2\delta^{A}_{C}\delta^{B}_{D}-\beta_3\delta^{A}_{D}\delta^{B}_{C},
\end{equation}
here, $\beta_i, i = 1, 2, 3$, dimensionless parameters, $\epsilon
= -1$ or $\epsilon = +1$ depending on whether the extra dimension
is time-like or space-like, respectively. Using the metric of the
bulk space as
\begin{equation}\label{}
dS^2 =
g_{\mu\nu}(x^\alpha,y)dx^{\mu}dx^{\nu}+\epsilon\phi(x^\alpha,y)dy^2,
\end{equation}
and
\begin{equation}\label{}
g_{AB}N^AN^{B} = \epsilon,\hspace{0.5cm}\epsilon^2 = 1,
\end{equation}
and assuming that the normal unit vector $N^A$ which is orthogonal
to the hypersurfaces $y = const.$ as, $N^A =
\frac{\delta^{A}_5}{\phi}$, the gravitational field equations on
the brane become as \cite{Ahmadi}
\begin{eqnarray}\label{GR}
G_{\mu\nu} &=& \frac{k_{5}^2}{2}
g_{\mu\nu}\Lambda_{5}-\frac{3(\epsilon+\alpha_1)}{(3+\alpha_1)}\left(K_{\mu\gamma}K_{\nu}^{\gamma}
-KK_{\mu\nu}
\right)-\frac{3(\epsilon+\alpha_3)}{2(3+\alpha_1)}g_{\mu\nu}K^2\nonumber\\
&+&\frac{3(\epsilon+\alpha_4)}{2(3+\alpha_1)}g_{\mu\nu}K_{\alpha\beta}K^{\alpha\beta}+
\left[\frac{\alpha_1(\alpha_5+\frac{1}{2})+3\alpha_5}{(3+\alpha_1)}\right]g_{\mu\nu}\frac{\phi^{\alpha}_{;\alpha}}{\phi}
-\frac{\alpha_5(5+\alpha_1)}{2(3+\alpha_1)}g_{\mu\nu}\frac{\phi_{,\alpha}\phi^{\alpha}_{,}}{\phi^2}\nonumber\\
&-&\frac{2\alpha_1}{(3+\alpha_1)}\frac{\phi_{;\nu\mu}}{\phi}
+\frac{4\alpha_5}{(3+\alpha_1)}\frac{\phi_{,\mu}\phi_{,\nu}}
{\phi^2}-\frac{3(\epsilon+\alpha_1)}{(3+\alpha_1)}{\cal
E_{\mu\nu}},
\end{eqnarray}
where
\begin{eqnarray}\label{}
\alpha_1 &=& 2(\beta_1+\beta_3),\hspace{0.5cm}\alpha_2 =
\frac{2\epsilon(\beta_1+\beta_2+\beta_3)}{3-2\epsilon(\beta_1+4\beta_2+\beta_3)},\nonumber\\\alpha_3
&=&
\frac{\alpha_1(3+\epsilon-2\beta_2)}{6}-\beta_2,\hspace{0.5cm}\alpha_4
= \frac{\alpha_1(6+\epsilon+\alpha_1)}{6},\hspace{0.5cm} \alpha_5
= \epsilon\beta_1.
\end{eqnarray}
The first partial derivatives in terms of the extrinsic curvature
of the brane is given by
\begin{equation}\label{}
K_{\mu\nu} = \frac{1}{2}{\cal L}_{N}g_{\mu\nu} =
\frac{1}{2\phi}\frac{\partial g_{\mu\nu}}{\partial
y},\hspace{0.5cm}K_{A5} = 0,
\end{equation}
and
\begin{equation}\label{}
{\cal E}_{\mu\nu} = C^{(5)}_{\mu A\nu B}N^AN^B.
\end{equation}
Using equation (\ref{GR}) and the $Z_2$ symmetry, we obtain the
gravitational field equations on the brane as
\begin{eqnarray}\label{23a}
G_{\mu\nu} &=& -\Lambda_4
g_{\mu\nu}+\kappa_4^2\tau_{\mu\nu}+\kappa_5^4\pi_{\mu\nu}-\frac{3(\epsilon+\alpha_1)}{(3+\alpha_1)}{\cal
E}_{\mu\nu}+{F}_{\mu\nu}+\left[\frac{\alpha_1(\alpha_5+\frac{1}{2})+3\alpha_5}{(3+\alpha_1)}\right]
g_{\mu\nu}\frac{\phi^{\alpha}_{;\alpha}}{\phi}\nonumber\\
&-&\frac{\alpha_5(5+\alpha_1)}{2(3+\alpha_1)}g_{\mu\nu}\frac{\phi_{,\alpha}\phi_{,}^{\alpha}}{\phi^2}-
\frac{2\alpha_1}{(3+\alpha_1)}\frac{\phi_{;\nu\mu}}{\phi}+\frac{4\alpha_5}{(3+\alpha_1)}\frac{\phi_{,\mu}\phi_{,\nu}}{\phi^2},
\end{eqnarray}
where
\begin{equation}\label{24a}
\Lambda_4 =
\frac{\kappa_5^2}{2}\Lambda_5+\kappa_5^4\left[\frac{\epsilon-3(-\alpha_1+4\alpha_6+16\alpha_7+
2\alpha_4)}{4(3+\alpha_1)(1+\epsilon\alpha_1)^2}\right]\lambda^2,
\end{equation}
\begin{equation}\label{25a}
\kappa_4^2 =
\kappa_5^4\left[\frac{2\epsilon-3(-2\alpha_1+4\alpha_6)}{4(3+\alpha_1)(1+\epsilon\alpha_1)^2}\right]\lambda,
\end{equation}
\begin{eqnarray}\label{26a}
\pi_{\mu\nu} &=&
\frac{3}{4(3+\alpha_1)(1+\epsilon\alpha_1)^2}\left[-(\epsilon+\alpha_1)
\tau_{\mu\alpha}\tau^{\alpha}_{\nu}+
(\frac{\epsilon}{3}+\alpha_6)\tau\tau_{\mu\nu}-(\frac{\epsilon}{6}-\alpha_7)g_{\mu\nu}\tau^2+
\frac{(\epsilon+\alpha_4)}{2}g_{\mu\nu}\tau_{\alpha\beta}\tau^{\alpha\beta}
\right]\nonumber\\&+&\left[\frac{3(\alpha_6+8\alpha_7+\alpha_4)}{4(3+\alpha_1)(1+\epsilon\alpha_1)^2}\right]g_{\mu\nu}\lambda\tau,
\end{eqnarray}
\begin{equation}\label{28a}
{F}_{\mu\nu} = \left[\frac{\kappa_5^2}{2}g_{\mu\nu}{\cal
T}_{\,\,\,\,y}^{y}+\frac{2\kappa_5^2}{3}\left({\cal
T}_{\mu\nu}-\frac{1}{4}g_{\mu\nu}{\cal
T}_{\,\,\,\,\alpha}^{\alpha}\right)\right]|_{y=0},
\end{equation}
and
\begin{equation}\label{19a}
\alpha_6 = \frac{\alpha_1(1-2\alpha_2)-2\epsilon\alpha_2}{3},
\end{equation}
\begin{equation}\label{20a}
\alpha_7 =
\frac{(\epsilon+\alpha_1)(\alpha_2+\alpha_2^2)}{3}+\frac{(\alpha_4-3\epsilon-4\alpha_3)(\alpha_2+2\alpha_2^2)}{9}-\frac{(\alpha_3+2\alpha_4)}{18}.
\end{equation}
For latter convenience we chose $\alpha_0 = 2\beta_2$, so that
constants $\alpha_i$ $(i = 2,...,7)$ are defined as a function of
$\alpha_0$ and $\alpha_1$. Now, by considering $N^{A}_{5} =
\delta^{A}_{5}$, $\epsilon = +1$ and ignoring the bulk matter,
$F_{\mu\nu} = 0$, we obtain the gravitational field equations in
four-dimensions as
\begin{equation}\label{23a}
G_{\mu\nu} = -\Lambda_4
g_{\mu\nu}+\kappa_4^2\left(\tau_{\mu\nu}+\frac{\alpha_1}{12}g_{\mu\nu}\tau\right)+\kappa_5^4\pi_{\mu\nu}-{
\tilde{\cal E}}_{\mu\nu},
\end{equation}
where
\begin{equation}\label{24a}
\Lambda_4 =
\frac{\kappa_5^2}{2}\Lambda_5+\frac{\kappa_5^4}{4(3-4\alpha_0-\alpha_1)}\lambda^2,
\end{equation}
\begin{equation}\label{25a}
\kappa_4^2 =
\frac{3\kappa_5^4}{2(3+\alpha_1)(3-4\alpha_0-\alpha_1)}\lambda,
\end{equation}
\begin{eqnarray}\label{26a}
\pi_{\mu\nu} &=&
\frac{3}{4(3+\alpha_1)(1+\alpha_1)}[\frac{(1-2\alpha_0-\alpha_1)}{(3-4\alpha_0-\alpha_1)}\tau\tau_{\mu\nu}-
\tau_{\,\,\,\,\mu}^{\alpha}\tau_{\alpha\nu}+\frac{(6+\alpha_1)}{12}g_{\mu\nu}\tau_{\alpha\beta}\tau^{\alpha\beta}\nonumber\\
&-&
\frac{2(3-\alpha_1)-(9+\alpha_1)\alpha_0}{12(3-4\alpha_0-\alpha_1)}g_{\mu\nu}\tau^2],
\end{eqnarray}
and
\begin{equation}\label{27a}
\tilde{{\cal E}}_{\mu\nu} =
\frac{3(1+\alpha_1)}{(3+\alpha_1)}{\cal E}_{\mu\nu}.
\end{equation}
We note that the model is different from the SMS model in two
cases. The first one is the existence of the trace part of the
brane energy-momentum tensor in the modified gravitational field equations on
the brane. This trace part of the energy-momentum tensor vanishes when
$\alpha_1 = 0$. The second departure from the SMS model arises
from definition of the fundamental quantities $\Lambda_4$ and
$\kappa_4^2$ which contain higher-dimensional modifications to the standard general relativity.

The Codazzi equation also implies the conservation of the
energy-momentum tensor of the matter on the brane
\begin{equation}\label{29a}
\nabla_{\mu}\tau^{\mu}_{\,\,\,\,\nu}+\alpha_2\nabla_{\nu}\tau-(1+4\alpha_2)\nabla_{\nu}\lambda
= -2(\epsilon+\alpha_1){\cal T}^{y}_{\,\,\,\,\nu},
\end{equation}
thus the brane energy-momentum tensor $\tau^{\mu}_{\,\,\,\,\nu}$
is not conserved. Moreover, the contracted Bianchi identities on
the brane imply that the projected Weyl tensor should obey the
constraint
\begin{equation}\label{30a}
\nabla_{\mu}{\tilde{\cal E}}^{\mu}_{\,\,\,\,\nu}=
-\nabla_{\nu}\Lambda_{4}+\kappa_4^2\left(\nabla_{\mu}\tau^{\mu}_{\,\,\,\,\nu}+\frac{\alpha_1}{12}\nabla_{\nu}\tau\right)
+\kappa_5^4\nabla_{\mu}\pi^{\mu}_{\,\,\,\,\nu}+\frac{\kappa_5^2}{2}\nabla_{\nu}{\cal
T}_{\,\,\,\,y}^{y}+\frac{2\kappa_5^2}{3}\left(\nabla_{\mu}{\cal
T}^{\mu}_{\,\,\,\,\nu}-\frac{1}{4}\nabla_{\nu}{\cal
T}^{\alpha}_{\,\,\,\,\alpha}\right).
\end{equation}
For convenience we suppose that the bulk cosmological constant and
brane tension is constant. Thus equations (\ref{29a}) and
(\ref{30a}) become
\begin{equation}\label{31a}
\nabla_{\mu}\tau^{\mu}_{\,\,\,\,\nu} = -\alpha_2\nabla_{\nu}\tau,
\end{equation}
\begin{equation}\label{32a}
\nabla_{\mu}{\tilde{\cal E}}^{\mu}_{\,\,\,\,\nu}=
-\kappa_4^2\left(\alpha_2-\frac{\alpha_1}{12}\right)\nabla_{\nu}\tau
+\kappa_5^4\nabla_{\mu}\pi^{\mu}_{\,\,\,\,\nu}.
\end{equation}
Equations (\ref{23a}), (\ref{31a}) and equation (\ref{32a}) give a
complete set of field equations for the brane gravitational field.
In the next section, we study the cosmological consequence of
anisotropic brane in the framework of such model.

\section{Bianchi type I brane cosmology}
In the following we will investigate the influence of the Lorentz
violating parameters on the anisotropic universe described by
Bianchi type I geometry. Considering $\beta_i = 0, i = 1,2,3$, we
can reduce the model to the SMS brane-world model and compare our
results to this model.

The line-element of a Bianchi type I space-time, which is a
generalization of the isotropic flat Friedmann-Robertson-Walker
(FRW) metric, is described by
\begin{eqnarray}\label{33a}
ds^2 = - dt \otimes dt + \sum_{i=1}^{3} a_i(t)^2 dx^i \otimes
dx^i,
\end{eqnarray}
where $a_i(t), i=1,2,3$ are the expansion factors in different
spatial directions. For later convenience we define the following
variables
\begin{eqnarray}\label{34a}
v = \prod_{i=1}^3 a_i,
\end{eqnarray}
\begin{eqnarray}\label{35a}
H_i = \frac{\dot a_i}{a_i},\hspace{0.5cm}i=1,2,3,
\end{eqnarray}
\begin{eqnarray}\label{36a}
3H = \sum_{i=1}^3 H_i.
\end{eqnarray}
In above equation, $v$ is the volume scale factor, $H_{i},
i=1,2,3$ are the directional Hubble parameters, and $H$ is the
mean Hubble parameter. The physical observable are the mean
anisotropy parameter $A$, the deceleration parameter $q$ and the
shear scalar $\sigma^2$ which are defined as
\begin{eqnarray}\label{39a}
A = \frac{1}{3}\sum_{i=1}^3 \left( \frac{\Delta H_i}{H} \right)^2,
\end{eqnarray}
\begin{eqnarray}\label{41a}
q = \frac{d}{d t}(\frac{1}{H})-1 = -\frac{1}{H^2}\left(\dot
H+H^2\right),
\end{eqnarray}
\begin{eqnarray}
\sigma^2 = \frac{1}{2}\sigma_{ij}\sigma^{ij} =
\frac{1}{2}\left(\sum_{i=1}^3 H_i^2-3H^2\right),
\end{eqnarray}
where $\Delta H_i = H_i - H$. The sign of the deceleration
parameter indicates how the universe expands. A positive sign for
$q$ corresponds to a decelerating universe whereas a negative sign
indicates inflation. We note that $A=0$ for an isotropic
expansion.

We also assume that the confined matter source on the brane is a
perfect fluid with equation of state $p=(\gamma-1)\rho$ where
$1\leq\gamma<2$.

Using the variables (\ref{34a})-(\ref{36a}), the Einstein
gravitational field equation (\ref{23a}), the Bianchi identity
(\ref{31a}) and the evolution equation for non-local dark
radiation (\ref{32a}) take the form
\begin{eqnarray}\label{42a}
3 \dot H + \sum_{i=1}^3 H_i^2 &=&
\Lambda_4-\frac{\kappa_4^2}{2}\left[(3\gamma-2)+\frac{\alpha_1}{6}(3\gamma-4)\right]\rho-
\frac{\kappa_5^4(\alpha_8+3\alpha_9)}{2}{\rho^2}
+\frac{3(1+\alpha_1)}{(3+\alpha_1)}{\cal U},
\end{eqnarray}
\begin{eqnarray}\label{43a}
\frac{1}{v} \frac{d}{d t} (v H_i) &=&
\Lambda_4-\frac{\kappa_4^2}{2}\left[(\gamma-2)+\frac{\alpha_1}{6}(3\gamma-4)\right]\rho
+\frac{\kappa_5^4(\alpha_8-\alpha_9)}{2}\rho^2
-\frac{(1+\alpha_1)}{(3+\alpha_1)}{\cal U},
\end{eqnarray}
and
\begin{equation}\label{44a}
\dot{\rho}\left[1+\alpha_2(4-3\gamma)\right]+3\gamma H\rho=0,
\end{equation}

\begin{equation}\label{45a}
\dot{{\cal U}}+{4}{H}{\cal U} =
\frac{\kappa_4^2\gamma(\alpha_2-\frac{\alpha_1}{12})(4-3\gamma)(3+\alpha_1)}{(1+\alpha_1)[1+\alpha_2(4-3\gamma)]}
H\rho-\frac{\kappa_5^4\alpha_{10}(3+\alpha_1)}{(1+\alpha_1)[1+\alpha_2(4-3\gamma)]}H\rho^2,
\end{equation}
where $3H = \frac{\dot{v}}{v}$ and constants $\alpha_8$,
$\alpha_9$ and $\alpha_{10}$ are defined as
$$
\alpha_8 =
\frac{-4(1+\alpha_1)(3+\alpha_1)+6\gamma(1+\alpha_1)(4+\alpha_1)-3\gamma^2(5+\alpha_1)(\alpha_0+\alpha_1)}
{16(3+\alpha_1)(1+\alpha_1)(3-4\alpha_0-\alpha_1)},
$$
$$
\alpha_9 =
\frac{4(1+\alpha_1)(3+\alpha_1)-6\gamma\alpha_1(1+\alpha_1)-3\gamma^2(\alpha_0+\alpha_1)(3-\alpha_1)}{16(3+\alpha_1)(1+\alpha_1)(3-4\alpha_0-\alpha_1)},
$$
$$
\alpha_{10} =
2\gamma\alpha_8+(\alpha_8+\alpha_9)[1+\alpha_2(4-3\gamma)].
$$
For $\gamma\neq0$ the solution of equation (\ref{44a}) is given by
\begin{equation}\label{46a}
\rho = \rho_0 v^{\frac{-\gamma}{1+(4-3\gamma)\alpha_2}}.
\end{equation}
Also integrating equation (\ref{45a}) yields
\begin{equation}\label{47a}
{\cal U} =  -3c
v^{-4/3}+\frac{\kappa_4^2\rho_0\gamma(\alpha_2-\frac{\alpha_1}{12})(4-3\gamma)(3+\alpha_1)}{[4+4\alpha_2(4-3\gamma)-3\gamma](1+\alpha_1)}
v^{\frac{-\gamma}{1+(4-3\gamma)\alpha_2}}-\frac{\kappa_5^4\rho_0^2\alpha_{10}(3+\alpha_1)}{[4+4(4-3\gamma)\alpha_2-6\gamma](1+\alpha_1)}
v^{\frac{-2\gamma}{1+(4-3\gamma)\alpha_2}},
\end{equation}
where $c$ is a constant of integration. Using the non-local energy
density and the evolution law of energy density, the gravitational
field equations (\ref{42a}) and (\ref{43a}) become
\begin{eqnarray}\label{48a}
3 \dot H + \sum_{i=1}^3 H_i^2 &=&
\Lambda_4-\frac{\kappa_4^2\rho_0}{2}\left[(3\gamma-2)+
\frac{\alpha_1}{6}(3\gamma-4)\right]v^{\frac{-\gamma}{1+(4-3\gamma)\alpha_2}}-
\frac{\kappa_5^4\rho_0^2}{2}(\alpha_8+3\alpha_9)v^{\frac{-2\gamma}{1+(4-3\gamma)\alpha_2}}\nonumber\\
&+&\frac{3(1+\alpha_1)}{(3+\alpha_1)}{\cal U}, \qquad i=1,2,3,
\end{eqnarray}
\begin{eqnarray}\label{49a}
\frac{1}{v} \frac{d}{d t} (v H_i) &=&
\Lambda_4-\frac{\kappa_4^2\rho_0}{2}\left[(\gamma-2)+\frac{\alpha_1}{6}(3\gamma-4)\right]
v^{\frac{-\gamma}{1+(4-3\gamma)\alpha_2}}
+\frac{\kappa_5^4\rho_0^2}{2}(\alpha_8-\alpha_9)v^{\frac{-2\gamma}{1+(4-3\gamma)\alpha_2}}\nonumber\\
&-&\frac{(1+\alpha_1)}{(3+\alpha_1)}{\cal U}, \qquad i=1,2,3.
\end{eqnarray}
Summing equations (\ref{49a}) we find
\begin{eqnarray}\label{50a}
\frac{1}{v} \frac{d}{d t} (3v H) &=&
3\Lambda_4-\frac{3\kappa_4^2}{2}\rho_0\left[(\gamma-2)+\frac{\alpha_1}{6}(3\gamma-4)\right]
v^{\frac{-\gamma}{1+(4-3\gamma)\alpha_2}}\nonumber\\
&+&\frac{3\kappa_5^4\rho_0^2}{2}(\alpha_8-\alpha_9)v^{\frac{-2\gamma}{1+(4-3\gamma)\alpha_2}}-\frac{3(1+\alpha_1)}{(3+\alpha_1)}{\cal
U}.
\end{eqnarray}
Now, substituting back equation (\ref{50a}) into equations
(\ref{49a}) we obtain
\begin{eqnarray}\label{51a}
H_i = H + \frac{h_i}{v}, \qquad i=1,2,3,
\end{eqnarray}
with $h_i, \, i=1,2,3$ being constants of integration satisfying
the consistency condition $\sum_{i=1}^3 h_i=0$. The basic equation
describing the dynamics of the anisotropic brane-world model with
Lorentz violation in the bulk can be written as
\begin{eqnarray}\label{52a}
\ddot v = 3\Lambda_4v+\frac{9c(1+\alpha_1)}{(3+\alpha_1)}v^{-1/3}+
f(\alpha_0,\alpha_1,\gamma)v^{\frac{(1-\gamma)+(4-3\gamma)\alpha_2}{1+(4-3\gamma)\alpha_2}}
+{\cal
F}(\alpha_0,\alpha_1,\gamma)v^{\frac{(1-2\gamma)+(4-3\gamma)\alpha_2}{1+(4-3\gamma)\alpha_2}},
\end{eqnarray}
where
$$
f(\alpha_0,\alpha_1,\gamma) =
\kappa_5^4\lambda\rho_0\frac{18(1+\alpha_1)[(2-\gamma)+\frac{\alpha_1}{6}(4-3\gamma)]-\gamma(12\alpha_0+9\alpha_1+
\alpha_1^2+4\alpha_0\alpha_1)}{8(1+\alpha_1)(3+\alpha_1)(3-4\alpha_0-\alpha_1)},
$$

$$
{\cal F}(\alpha_0,\alpha_1,\gamma) =
\kappa_5^4\rho_0^2\left[\frac{3[4(3+\alpha_1)+3\gamma^2(\alpha_0+\alpha_1)-6\gamma(2+\alpha_1)]}{16(3+\alpha_1)
(3-4\alpha_0-\alpha_1)}+
\frac{\alpha_{10}(3-4\alpha_0-\alpha_1)}{4(1+\alpha_1)+2\gamma(2\alpha_0-\alpha_1-3)}\right].
$$
The general solution of equation (\ref{52a}) becomes
\begin{eqnarray}\label{53a}
t - t_0 &=& \int (3\Lambda_4
v^2+\frac{27c(1+\alpha_1)}{(3+\alpha_1)}v^{2/3}+2g(\alpha_0,\alpha_1,\gamma)v^{\frac{(2-\gamma)+2(4-3\gamma)\alpha_2}
{1+(4-3\gamma)\alpha_2}} \nonumber\\
&+& 2{\cal G}
(\alpha_0,\alpha_1,\gamma)v^{\frac{2(1-\gamma)+2(4-3\gamma)\alpha_2}{1+(4-3\gamma)\alpha_2}}+{
C})^{-1/2}dv,
\end{eqnarray}
where
$$
g(\alpha_0,\alpha_1,\gamma)=\frac{1+(4-3\gamma)\alpha_2}{(2-\gamma)+2(4-3\gamma)\alpha_2}f(\alpha_0,\alpha_1,\gamma),
$$

$$
{\cal
G}(\alpha_0,\alpha_1,\gamma)=\frac{1+(4-3\gamma)\alpha_2}{2(1-\gamma)+2(4-3\gamma)\alpha_2}{\cal
F}(\alpha_0,\alpha_1,\gamma),
$$
where $C$ is a constant of integration. The time variation of the
physically important parameters described above in the exact
parametric form, with $v$ taken as a parameter, is given by
\begin{eqnarray}\label{56a}
A = 3h^2\left[3\Lambda_4
v^2+\frac{27c(1+\alpha_1)}{(3+\alpha_1)}v^{2/3}+2g(\alpha_0,\alpha_1,\gamma)v^{\frac{(2-\gamma)
+2(4-3\gamma)\alpha_2}{1+(4-3\gamma)\alpha_2}} +2{\cal G}
(\alpha_0,\alpha_1,\gamma)v^{\frac{2(1-\gamma)+2(4-3\gamma)\alpha_2}{1+(4-3\gamma)\alpha_2}}+{
C}\right]^{-1},
\end{eqnarray}
\begin{eqnarray}\label{57a}
q =
2-\frac{3v\left[3\Lambda_4v+\frac{9c(1+\alpha_1)}{(3+\alpha_1)}v^{-1/3}+
f(\alpha_0,\alpha_1,\gamma)v^{\frac{(1-\gamma)+(4-3\gamma)\alpha_2}{1+(4-3\gamma)\alpha_2}}
+{\cal
F}(\alpha_0,\alpha_1,\gamma)v^{\frac{(1-2\gamma)+(4-3\gamma)\alpha_2}{1+(4-3\gamma)\alpha_2}}\right]}{3\Lambda_4
v^2+\frac{27c(1+\alpha_1)}{(3+\alpha_1)}v^{2/3}+2g(\alpha_0,\alpha_1,\gamma)v^{\frac{(2-\gamma)+2(4-3\gamma)\alpha_2}{1+(4-3\gamma)\alpha_2}}
+2{\cal G}
(\alpha_0,\alpha_1,\gamma)v^{\frac{2(1-\gamma)+2(4-3\gamma)\alpha_2}{1+(4-3\gamma)\alpha_2}}+{
C}},
\end{eqnarray}
\begin{eqnarray}\label{58a}
a_i &=& a_{0i} v^{1/3}\exp\{\int[3\Lambda_4
v^4+\frac{27c(1+\alpha_1)}{(3+\alpha_1)}v^{2/3}+2g(\alpha_0,\alpha_1,\gamma)v^{\frac{(2-\gamma)+
2(4-3\gamma)\alpha_2}{1+(4-3\gamma)\alpha_2}}\nonumber\\
&+& 2{\cal G}
(\alpha_0,\alpha_1,\gamma)v^{\frac{2(1-\gamma)+2(4-3\gamma)\alpha_2}{1+(4-3\gamma)\alpha_2}}+{C}v^2]^{-1/2}dv\},\hspace{0.5
cm} i=1,2,3,
\end{eqnarray}
\begin{eqnarray}\label{60a}
\sigma^2 = \frac{h^2}{2 v^2},
\end{eqnarray}
where $h^2=\sum_{i=1}^3 h_i^2$. Also, the integration constants
$h_{i}$ and ${C}$ must satisfy the consistency condition
$h^2=\frac{2}{3}{C}$. As one can see from equations
(\ref{56a})-(\ref{58a}) the behavior of these physical parameters
depends on the Lorentz violating parameters and the equation of
state of the cosmological fluid.

For a well-defined theory, the following constraints put on the
parameters $\beta_i$ as \cite{32}
$$
\beta_1\geq0,
$$
$$
(\beta_1+\beta_2+\beta_3)/\beta_1\leq1,
$$
$$
(\beta_1+\beta_2+\beta_3)/\beta_1\geq0,
$$
$$
\beta_1+\beta_3\leq0.
$$
\vspace{1.2mm}\noindent\\
\begin{center}
\begin{tabular}{cccccc} \hline\hline
\multicolumn{2}{c}                 {constraints for $G>0$} & $G$ & constraints of ref \cite{32}& integer values of $\alpha_0$ and $\alpha_1$  \\
                                   & &  &   &
                                  which satisfy both of two
                                   constraints
\\ \hline \hline
$\lambda>0$       & $3+\alpha_1>0$    &    positive &$\alpha_0\geq0$ & $\alpha_0=0,$ $\alpha_1=0, 1, 2$ \\
                  &$3-4\alpha_0-\alpha_1>0$    &              &$\alpha_0+\alpha_1\geq0$ &         \\
\cline{2-5}
            & $3+\alpha_1<0$    &    positive &$\alpha_0\geq0$ &  $\alpha_0 = \{4, 5, 6, ...\}$\\
                  &$3-4\alpha_0-\alpha_1<0$    &            &$\alpha_0+\alpha_1\geq0$  &$\alpha_1 = \{-\alpha_0, -\alpha_0+1, -\alpha_0+2...,-4\}$         \\
\hline
$\lambda<0$       & $3+\alpha_1<0$    &    positive &$\alpha_0\geq0$ & no values\\
                   &$3-4\alpha_0-\alpha_1>0$    &              &$\alpha_0+\alpha_1\geq0$ &         \\
\cline{2-5}
                  & $3+\alpha_1>0$    &    positive &$\alpha_0\geq0$ & $\alpha_0=0,$ $\alpha_1>3$\\
                  &$3-4\alpha_0-\alpha_1<0$    &              &$\alpha_0+\alpha_1\geq0$ &  $\alpha_0 = 1,$ $\alpha_1>-1$        \\
                  &                        &               &                             &  $\alpha_0 = \{2, 3, 4, ...\},$
                  $\alpha_1>-3$\\
 \hline\hline
\end{tabular}\vspace{2mm}
\begin{center}
{\footnotesize Table 1: The possible values of Lorentz violating
parameters in a brane-world universe.}
\end{center}
\end{center}
\vspace{1.2mm}\noindent\\
The first condition results from the need for a positive-definite
Hamiltonian for the perturbations; the next two from demanding the
subluminal and non-tachyonic propagation of the spin-0 field; and
the last condition from insisting subluminal propagation of the
spin-2 field. Together these conditions imply that
$$
\beta_1+\beta_2+\beta_3\geq0,
$$
$$
\beta_2\geq0,
$$
which in terms of $\alpha_0$ and $\alpha_1$ we have
$$
\alpha_0+\alpha_1\geq0,
$$
$$
\alpha_0\geq0.
$$
The above conditions together with constraints require for a
positive effective Newtonian constant (\ref{25a}), lead to the
particular values of $\alpha_0$ and $\alpha_1$. We have summarized
these results in table 1.

The singular state at $t=0$ is characterized by the condition
$v(0) = 0$. The value of the anisotropy parameter for $t = 0$
depends on the Lorentz violating parameters and the equation of
the state. Hence for $\gamma=1$ equation (\ref{56a}) reduces to
\begin{equation}\label{27a}
A = \frac{3h^2}{3\Lambda_4 v^2
+\frac{27c(1+\alpha_1)}{(3+\alpha_1)}v^{\frac{2}{3}}+2g(\alpha_0,
\alpha_1,1)v^{\frac{1+2\alpha_2}{1+\alpha_2}}+2{\cal
G}(\alpha_0,\alpha_1,1)v^\frac{2\alpha_2}{1+\alpha_2}+{C}},
\end{equation}
the singular behavior of the dust filled brane universe depends on
$\alpha_2$, so that from above equation it follows
\begin{equation}\label{27a}
\lim _{v\rightarrow 0} A(v) = 0,\hspace{0.5 cm}-1<\alpha_2<0,
\end{equation}
and
\begin{equation}\label{27a}
\lim _{v\rightarrow 0} A(v) =
\frac{36h^2}{k_5^{4}\rho_0^{2}+12{C}},\hspace{0.5 cm}\alpha_2 = 0,
\end{equation}
and otherwise
\begin{equation}\label{27a}
\lim _{v\rightarrow 0}A(v) = \frac{3h^2}{{C}}.
\end{equation}
In the absence of the Lorentz violating parameters the dust filled
brane universes are born in an anisotropic state \cite{Mak}, while
in the brane universe with the Lorentz invariance violation the
early time behavior of the brane universe is sensitive to the
values of $\alpha_2$, so that it is possible to admitting of both
the isotropic and anisotropic state. Therefore the singular state
of the dust filled Bianchi type I brane universe in our model is
different from the case of brane cosmological models without the
bulk Lorentz violation.

For the case of the radiation anisotropic brane universe, $p
 = \frac{1}{3}\rho$ and $\gamma=\frac{4}{3}$, equation (\ref{56a}) reduces to
\begin{equation}\label{above}
A = \frac{3h^2}{3\Lambda_4 v^2
+\frac{27c(1+\alpha_1)}{(3+\alpha_1)}v^{\frac{2}{3}}+2g(\alpha_0,
\alpha_1,4/3)v^{\frac{2}{3}}+2{\cal
G}(\alpha_0,\alpha_1,4/3)v^{-\frac{2}{3}}+{C}},
\end{equation}
in this case the singular behavior of the anisotropy parameter is
independent of $\alpha_2$, so that from equation (\ref{above}), we
have
\begin{equation}\label{27a}
\lim _{v\rightarrow 0} A(v) = 0.
\end{equation}
In figure 1, we have plotted the anisotropy parameter for
$\alpha_0 = 0$ and $\alpha_1 = 2$. Since $\alpha_2$ expressed in
terms of $\alpha_0$ and $\alpha_1$ becomes $\alpha_2 =
\frac{\alpha_0+\alpha_1}{3-4\alpha_0-\alpha_1}$, therefore
$\alpha_2 = 2$ and from equation (52) for $\gamma = 1$ we have
$A(0) = \frac{3h^2}{{C}}$ and the anisotropic brane universe
starts its evolution from an anisotropic state to an isotropic de
Sitter inflationary phase at late time. For $\gamma = 4/3$, the
late time behavior of anisotropy parameter is isotropic with $A(0)
= 0$. The behavior of the deceleration parameter of the Bianchi
type I geometry is illustrated, for different values of $\gamma$,
$\alpha_0 = 0$ and $\alpha_1 = 2$, in figure 2. In the initial stage the behavior of the
Bianchi type I brane universe is non-inflationary,
but brane universe ends in an accelerated expanding stage at late
time. The time variation of the shear
parameter as a function of time is represented in figure 3. The
shear is a decreasing function of time and in the limit of large
time corresponding to the isotropic limit, $\sigma^2\rightarrow0$.
In the limit of small time the shear has a singular behavior
tending to infinity, $\sigma^2\rightarrow\infty$.

For $\beta_2 = 0$ and $\beta_1 = -\beta_3$, equations
(\ref{56a})-(\ref{58a}) give the general solutions of the physical
parameters on the volume scale factor for the generalized RS II
model \cite{Harko}
\begin{equation}\label{}
a_i = a_{0i}v^{1/3} \exp\left[h_{i}\int\left(3\Lambda_4
v^4+3\kappa_4^2\rho_0v^{4-\gamma}+\frac{1}{4}\kappa_5^4\rho_0^2v^{4-2\gamma}+Cv^2\right)^{-1/2}dv\right],\hspace{0.5
cm} i=1,2,3,
\end{equation}
\begin{eqnarray}\label{61a}
q=2-\frac{36\Lambda_4
v^2+18(2-\gamma)\kappa_4^2\rho_0v^{2-\gamma}+3(1-\gamma)\kappa_5^4\rho_0^2v^{2-2\gamma}}{12\Lambda_4
v^2+12\kappa_4^2\rho_0v^{2-\gamma}+\kappa_5^4\rho_0^2v^{2-2\gamma}+4C},
\end{eqnarray}
\begin{eqnarray}\label{60a}
A = 3h^2\left(3\Lambda_4
v^2+3\kappa_4^2\rho_0v^{2-\gamma}+\frac{1}{4}\kappa_5^4\rho_0^2v^{2-2\gamma}+C\right)^{-1},
\end{eqnarray}
\begin{eqnarray}\label{60a}
\sigma^2 = \frac{h^2}{2 v^2},
\end{eqnarray}

where $h^2=\sum_{i=1}^{3}h_i^2$ and $C$ is a constant of
integration. The behavior of the deceleration parameter of the
Bianchi type I geometry is illustrated, for different values of
$\gamma$, in figure 4. In the initial stage the behavior of the
Bianchi type I brane universe is non-inflationary, but brane
universe ends in an accelerated expanding stage at late time.

In figure 5, we have also plotted the anisotropy parameter for
different values of $\gamma$. The behavior of the anisotropy
parameter shows that at high densities the brane universe starts
its evolutions from an isotropic state with $A(0)=0$ for $\gamma =
4/3$, and ends in an isotropic de Sitter inflationary phase at
late time. An important difference between the anisotropic
homogeneous brane-world cosmological models and the standard
general relativity is that brane universes are born in an
isotropic state.

The time variation of the shear parameter as a function of time is
represented in figure 6. The shear is a decreasing function of
time and in the limit of large time corresponding to the isotropic
limit, $\sigma^2\rightarrow0$. In the limit of small time the
shear has a singular behavior tending to infinity,
$\sigma^2\rightarrow\infty$.

\vspace{10mm}

\section{Conclusions}

In this letter, we have dealt with the dark energy problem in the
context of the brane-world scenario with the bulk Lorentz
violation, introduced by specifying a preferred frame through the
introduction of a dynamical vector field normal to the brane. In
our model, due to the local Lorentz violation in the bulk, the
Friedmann equations on the anisotropic brane have been modified by
the Lorentz violating parameters $\beta_i, i = 1, 2, 3$ and the
equation of state of the matter. Therefore, for a fixed value of
$\gamma$, the behavior of an anisotropic brane universe is
controlled by the Lorentz violating parameters. The behavior of
the anisotropy parameter shows that the brane universe evolves
from an isotropic/anisotropic state to an isotropic state which
has entered an accelerated expanding phase. The early time
behavior of anisotropic parameter is different from those obtained
in RS type II brane-world models without the bulk Lorentz
violation \cite{Harko}.

\begin{figure}
\centerline{\begin{tabular}{ccc}
\epsfig{figure=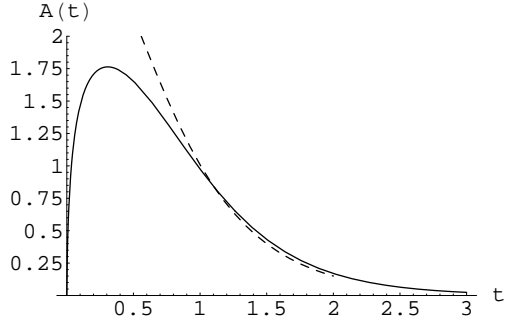,width=7cm}
\end{tabular} } \caption{\footnotesize The anisotropy parameter of the Bianchi type I brane geometry, with
the bulk Lorentz violation $\alpha_0 = 0$, $\alpha_1 = 2$, for
$\gamma = 4/3$ (solid curve) and for
 $\gamma = 1$ (dashed curve).}\label{figure1}
\end{figure}

\begin{figure}
\centerline{\begin{tabular}{ccc}\epsfig{figure=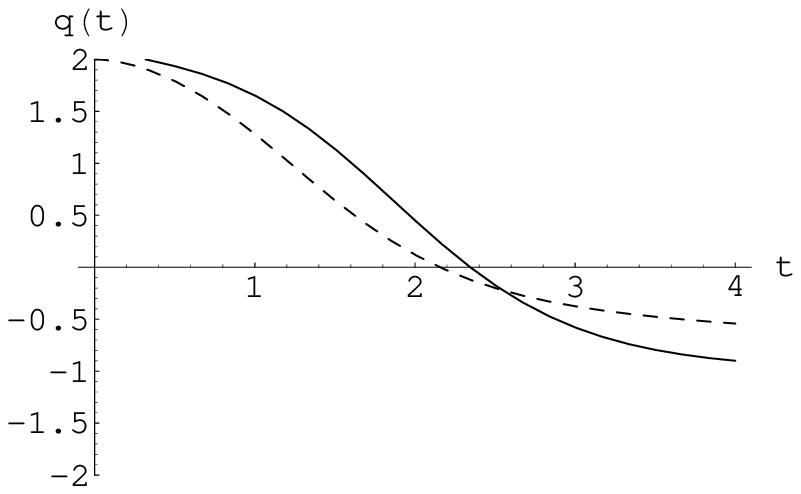,width=7cm}
\end{tabular} } \caption{\footnotesize The deceleration parameter of the Bianchi type I brane geometry, with
the bulk Lorentz violation $\alpha_0 = 0$, $\alpha_1 = 2$, for
 $\gamma = 4/3$ (solid curve) and for
 $\gamma = 1$ (dashed curve).}\label{figure1}
\end{figure}

\begin{figure}
\centerline{\begin{tabular}{ccc}
\epsfig{figure=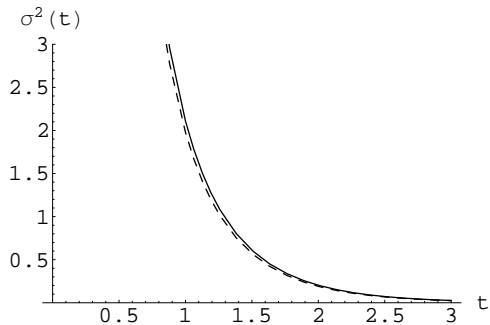,width=7cm}
\end{tabular} } \caption{\footnotesize  The shear parameter of the Bianchi type I brane universe, with
the bulk Lorentz violation $\alpha_0 = 0$, $\alpha_1 = 2$, for
 $\gamma = 4/3$ (solid curve) and for
 $\gamma = 1$ (dashed curve).}\label{figure1}
\end{figure}

\begin{figure}
\centerline{\begin{tabular}{ccc}
\epsfig{figure=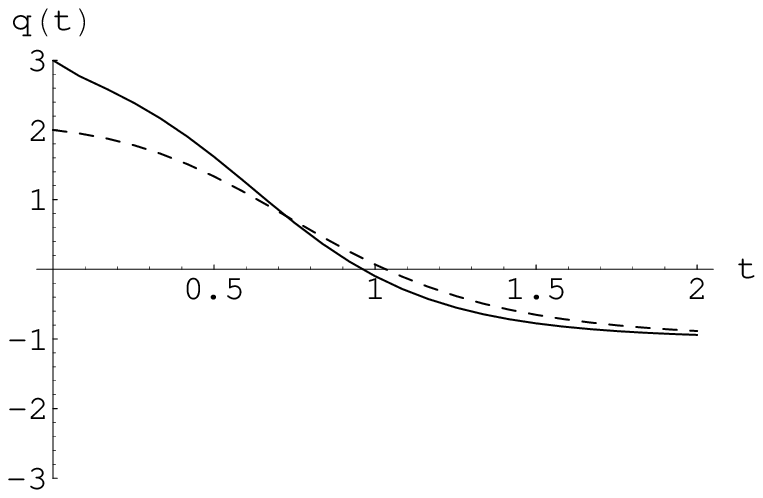,width=7cm}
\end{tabular} } \caption{\footnotesize  The deceleration parameter of the Bianchi type I brane universe with confined
perfect cosmological fluid for $\gamma = 4/3$ (solid curve) and
 $\gamma = 1$ (dashed curve).}\label{figure1}
\end{figure}

\begin{figure}
\centerline{\begin{tabular}{ccc}
\epsfig{figure=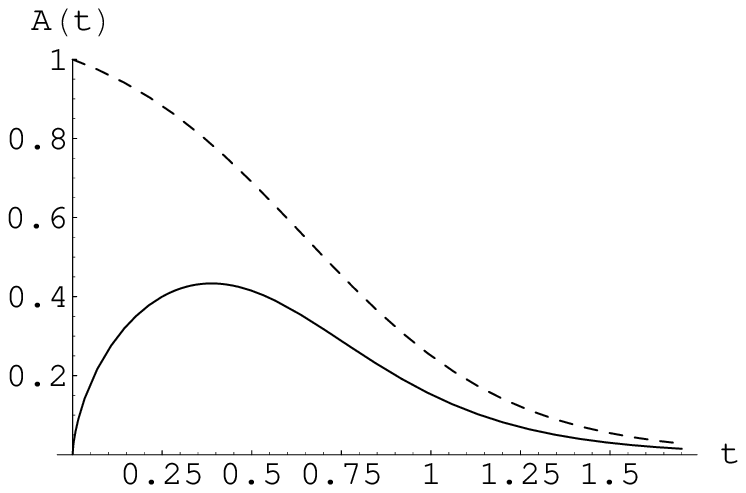,width=7cm}
\end{tabular} } \caption{\footnotesize  The anisotropy parameter of the Bianchi type I brane universe with confined
perfect cosmological fluid for $\gamma = 4/3$ (solid curve) and
 $\gamma = 1$ (dashed curve).}\label{figure1}
\end{figure}

\begin{figure}
\centerline{\begin{tabular}{ccc}
\epsfig{figure=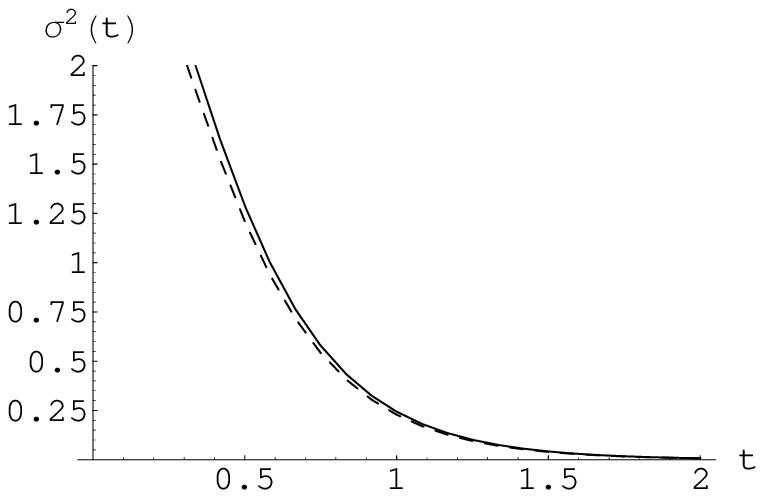,width=7cm}
\end{tabular} } \caption{\footnotesize  The shear parameter of the Bianchi type I brane universe with confined
perfect cosmological fluid for $\gamma = 4/3$ (solid curve) and
 $\gamma = 1$ (dashed curve).}\label{figure1}
\end{figure}

\end{document}